\documentclass[11pt]{article}  
\usepackage{menuproc}
\usepackage{cite}
\usepackage{epsfig}
\usepackage{amsmath,amssymb}
\begin{document}
%
%
%
\titlematter{Pion-nucleon sigma-term - a review}%
{M.E. Sainio}%
{Helsinki Institute of Physics, and\\
 Department of Physics, University of Helsinki, P.O. Box 64, 00014 Helsinki, 
 Finland}%
{A brief review of the pion-nucleon sigma-term is given. Aspects of both
chiral perturbation theory and phenomenology are discussed.}
%
%

\section{Introduction}

The pion-nucleon sigma-term is defined as
\begin{eqnarray}
\sigma = \frac{\hat{m}}{2 m_p} \langle p | \bar{u}u+\bar{d}d|p \rangle,
\hspace{1.cm} \hat{m}=\frac{1}{2}(m_u+m_d), \nonumber
\end{eqnarray}
i.e. as the proton matrix element of the $u$- and $d$- quark mass term of the
QCD hamiltonian ($m_p$ is the mass of the proton). More generally, sigma-terms are 
proportional to the scalar quark currents
\begin{eqnarray}
\langle A|m_q \bar{q}q|A \rangle \; \; ; q=u, d, s \; \; ; \ A=\pi, K, N. \nonumber
\end{eqnarray} 
These are of interest, because they are related to the hadron mass spectrum, to the
scattering amplitudes through Ward identities, to the strangeness content of $A$,
to the quark mass ratios and to the question of dark matter. For an early review of
the topic, see ref. \cite{reya}.

The pion-nucleon sigma-term is the $t=0$ value of the scalar form factor
\begin{eqnarray}
\bar{u}'\sigma(t) u = \hat{m}\, \langle p' |\bar{u}u+\bar{d}d|p \rangle, \; \; 
t=(p'-p)^2, \nonumber
\end{eqnarray}
i.e. $\sigma = \sigma(t=0)$.
The strangeness content of the proton can then be defined as
\begin{eqnarray}
y = \frac{2 \, \langle p | \bar{s}s |p \rangle}{\langle p | \bar{u}u +
\bar{d}d |p \rangle} \nonumber
\end{eqnarray}
(the OZI rule would imply $y$=0).

Algebraically the $\sigma$ can be written in the form
\begin{eqnarray}
\sigma = \frac{\hat{m}}{2 m_p} \frac{\langle p | \bar{u}u+\bar{d}d-2\bar{s}s|p \rangle}{1-y}, 
\nonumber
\end{eqnarray}
where the numerator is proportional to the octet breaking piece in the hamiltonian.
To first order in SU(3) breaking we have now
\begin{eqnarray}
\sigma \simeq \frac{\hat{m}}{m_s-\hat{m}} \frac{m_\Xi+m_\Sigma-2m_N}{1-y} \simeq 
\frac{26 \; {\rm MeV}}{1-y}, \nonumber
\end{eqnarray}
where the quark mass ratio
\begin{eqnarray}
\frac{m_s}{\hat{m}} = 2 \, \frac{M_K^2}{M^2_\pi}-1 \simeq 25 \nonumber
\end{eqnarray}
has been used.

Chiral perturbation theory (ChPT) allows for the determination of the combination
\begin{eqnarray}
\hat{\sigma} = \sigma (1-y) \nonumber
\end{eqnarray}
from the baryon spectrum. Therefore, if the sigma-term can be determined from data,
the \newline strangeness content $y$ can be estimated.

Section 2 will be dealing with the ChPT aspects of the sigma. The phenomenological
discussion will follow in section 3. A brief summary is given is section 4 together
with reference to recent developments in the lattice frontier.

\section{The $\sigma$ -term}

ChPT gives in leading order
\begin{eqnarray}
\hat{\sigma} \simeq 26  \; {\rm MeV} \nonumber 
\end{eqnarray}
as indicated above. The ${\cal O}(m_q^{3/2})$ calculation
of Gasser and Leutwyler \cite{g1,gl} yields
\begin{eqnarray}
\hat{\sigma} = 35 \pm 5 \; {\rm MeV.} \nonumber 
\end{eqnarray}
Borasoy and Mei\ss ner \cite{bm} have made the calculation  in the
heavy baryon framework of ChPT to order
${\cal O}(m_q^2)$ with the result
\begin{eqnarray}
\hat{\sigma} = 36 \pm 7  \; {\rm MeV.} \nonumber
\end{eqnarray}

\subsection{Scalar form factor}

Contact to pion-nucleon scattering can be made
at the unphysical Cheng-Dashen point ($s=u=m_N^2, t=2 M_\pi^2$)
and, therefore, it is of interest to determine the difference
of the scalar form factor
\begin{eqnarray}
\Delta_\sigma \equiv \sigma(2 M_\pi^2) - \sigma(0). \nonumber
\end{eqnarray} 
In leading order we have \cite{pp,gl}
\begin{eqnarray}
\Delta_\sigma = \frac{3 g_A^2 M_\pi^3}{64 \pi F_\pi^2} + {\cal O}(M_\pi^4 \ln M_\pi^2),
\nonumber
\end{eqnarray}
which is numerically about 7 MeV. ChPT to one loop yields $\Delta_\sigma \simeq 5$ MeV 
\cite{gss}. In heavy baryon ChPT (HBChPT) including the ${\cal O}(p^4)$ pieces due to 
the low-lying spin-3/2
baryons the result $\Delta_\sigma \simeq 15$ MeV is obtained \cite{bkm}.
A dispersion analysis where particular emphasis was on the treatment of
the $\pi \pi$ interaction dominating the curvature of the $\sigma(t)$ yields \cite{gls1}
\begin{eqnarray}
\Delta_\sigma = 15.2 \pm 0.4 \; {\rm MeV.} \nonumber
\end{eqnarray}
More recently, Becher and Leutwyler have calculated \cite{bl1} the scalar form
factor to order $p^4$ in a formulation of the baryon ChPT which
keeps the Lorentz and chiral invariance explicit at all stages. The result
for $\Delta_\sigma$ is
\begin{eqnarray}
\Delta_\sigma = 14.0 \; {\rm MeV} + 2 M^4 \bar{e}_2, \nonumber
\end{eqnarray}
where $M$ is the leading order result for $M_\pi$ $(M^2=2 \hat{m} B)$ and $\bar{e}_2$
is a renormalized coupling constant due to the ${\cal L}_N^{(4)}$
lagrangian. Comparison with the result of the dispersive calculation
shows that the piece proportional to $\bar{e}_2$ is small as it
should be.

The value of the form factor at $t=0$, i.e. the $\sigma$, can be
calculated from the quark mass expansion of the nucleon mass
by making use of the Feynman-Hellmann theorem
\begin{eqnarray}
\sigma=\hat{m} \frac{\partial m_N}{\partial \hat{m}} \nonumber
\end{eqnarray}
or equivalently 
\begin{eqnarray}
\sigma = M^2 \frac{\partial m_N}{\partial M^2}. \nonumber
\end{eqnarray}
The physical mass of the nucleon to order $p^4$ is \cite{bl1,l}
\begin{eqnarray}
m_N=m_0+k_1 M^2+k_2 M^3+k_3 M^4 \ln \frac{M^2}{m_0^2}
 +k_4 M^4+{\cal O}(M^5), \nonumber
\end{eqnarray}
where $m_0$ is the nucleon mass in the chiral limit and the
factors $k_i$ contain the low-energy constants.
This yields for $\sigma$
\begin{eqnarray}
\sigma = k_1 M^2 + \frac{3}{2} k_2 M^3 
+k_3 M^4 \{ 2 \ln \frac{M^2}{m_0^2}+1\} 
+2 k_4 M^4 + {\cal O}(M^5) \nonumber
\end{eqnarray}
and numerically
\begin{eqnarray}
\sigma = (75-23-7+0) \; {\rm MeV} = 45 \; {\rm MeV}, \nonumber
\end{eqnarray}
where the leading term, 75 MeV, is fixed by requiring the result
$\sigma = 45$ MeV for the sigma-term \cite{gls2}. Also, the
term $k_4 M^4$ is put equal to 0, because it is expected to be
very small.

A ${\cal O}(p^3)$ calculation in HBChPT, where the low-energy constants
are fixed inside the Mandelstam triangle, gives \cite{pbm} the value
$\sigma = 40$ MeV, if Karlsruhe phase shifts (KA84) are used as input.
The VPI input (SP99) would yield $\sigma \simeq 200$ MeV.

\subsection{Cheng-Dashen point}

As mentioned earlier, contact to the pion-nucleon interaction can
be made at the Cheng-Dashen point: A low-energy theorem of
chiral symmetry states
\begin{eqnarray}
\Sigma \equiv F^2_\pi \bar{D}^+(\nu=0, t=2M_\pi^2) = \sigma(2 M_\pi^2)+\Delta_R,
\nonumber
\end{eqnarray}
where $\nu=(s-u)/4m_N$, $\bar{D}^+$ is the isoscalar $D$-amplitude with the
pseudovector Born term subtracted and $\Delta_R$ is the remainder. The quantity
$\Delta_R$ is formally of the order $M_\pi^4$, and the one-loop
result [a ${\cal O}(p^3)$ value] is $\Delta_R = 0.35$ MeV \cite{gss}.
In HBChPT it has been shown that no logarithmic contribution to order $M_\pi^4$ 
appears \cite{bkm2}.
This result is verified in the ${\cal O}(p^4)$ calculation of the pion-nucleon
amplitude \cite{bl2}. Numerically, with the low-energy constants estimated with the
resonance exchange saturation, the result is $\Delta_R \simeq 2$ MeV \cite{bkm2}
which is considered to be the upper limit for $\Delta_R$. Therefore, it can
well be approximated as
\begin{eqnarray}
\Sigma \simeq \sigma(2 M^2_\pi). \nonumber
\end{eqnarray}
One may, of course, ask how this result would change, if the fact
$m_u-m_d \neq 0$ would be taken into account.

\section{$\Sigma$ phenomenology}

The standard expression for the $\pi$N amplitude is
\begin{eqnarray}
T_{\pi N} = \bar{u}'[A(\nu,t) + \frac{1}{2} \gamma^{\mu} (q+q')_{\mu}
            B(\nu,t)]u, \nonumber
\end{eqnarray}
where $q$ and $q'$ are the initial and final pion momentum respectively.
The $D$-amplitude is
\begin{eqnarray}
D(\nu,t) = A(\nu,t) + \nu B(\nu,t) \nonumber
\end{eqnarray}
and, through the optical theorem,
\begin{eqnarray}
{\rm Im}\, D(\omega,t=0) = k_{\, \rm lab} \, \sigma. \nonumber
\end{eqnarray}
Its imaginary part in the forward direction is directly fixed
by the cross section data ($\omega$ is the initial pion laboratory
energy). The isospin components are simply related to the amplitudes
in the particle basis
\begin{eqnarray}
D^{\pm} = \frac{1}{2} (D_{\pi^-p} \pm D_{\pi^+p}).  \nonumber
\end{eqnarray}
The relevant combination for the $\Sigma$ -term discussion
is the isoscalar piece, $D^+$, at the Cheng-Dashen point.

The standard value with the Karlsruhe input has been the result \cite{koch}
\begin{eqnarray}
\Sigma = 64 \pm 8 \; {\rm MeV} \nonumber
\end{eqnarray}
based on hyperbolic dispersion relations. The error reflects the internal
consistency of the method. An attempt to include an estimate of the error
in $\Sigma$ generated by the errors of the low-energy data was published in 
ref. \cite{gls2}. The numerical result there was $\Sigma \simeq 60$ MeV with
the Karlsruhe input.

The $\Sigma$ can also be related to the threshold parameters \cite{gasser}
\begin{eqnarray}
\Sigma=F_\pi^2 [L(a^+_{l\pm},\tau) + (1+\frac{M_\pi}{m_N})\tau J^+] + \delta^{ChPT},
\nonumber
\end{eqnarray}
where  $L$ is a linear combination of the threshold parameters and
$\tau$ is a free parameter, $J^+$ is the integral over the total cross section
\begin{eqnarray}
J^+ = \frac{2 M_\pi^2}{\pi} \, \int^\infty_0 \frac{\sigma^+(k')}{\omega(k')^2} \, dk' 
\nonumber
\end{eqnarray}
and $\delta^{ChPT}$ is the remainder from ChPT, see also
ref. \cite{ls}, where references to earlier work in a similar spirit can be found. 
In such a formulation the contribution from $a^+_{1+}$ to $\Sigma$ 
may vary from -150 MeV to 250 MeV for $\tau \in [-1,1]$.
Olsson has recently \cite{olsson} written a sum rule for $\Sigma$ which includes
an expansion in terms of threshold parameters. With the Karlsruhe input the consistent
result, $\Sigma = 55 \pm 6$ MeV, follows. With input from ref. \cite{gashi} the
value $\Sigma = 71 \pm 9$ MeV is obtained.

\subsection{Low-energy analysis}

At low energy the pion-nucleon interaction is dominated by six partial
waves, 2 $s$-waves and 4 $p$-waves. Therefore, six relations are needed
to pin down the six partial waves. Such relations can be obtained by
writing six dispersion relations for the $D^\pm$, $B^\pm$ and $E^\pm$
where
\begin{eqnarray}
E^{\pm} = \frac{\partial}{\partial t} (A^{\pm} + \omega B^{\pm})|_{t=0}.
\nonumber
\end{eqnarray}
There are two subtraction constants in the 6 dispersion relations,
one for $D^+$ and $E^+$  ($x=M_\pi/m_N$):
\begin{eqnarray}
\bar{D}^{+}(\mu) &=& 4\pi(1+x)a^{+}_{0+} + {g^{2}x^{3} \over
M_\pi(4-x^{2})} \nonumber \\
\bar{E}^{+}(\mu) &=& 6\pi(1+x)a^{+}_{1+} - {g^2 x^2 \over
M_\pi^{3}(2-x)^2}. \nonumber
\end{eqnarray}
As described in ref. \cite{gls2} the six dispersion relations can be
solved iteratively with input for the invariant amplitudes from high energy 
(here $k_{lab} \geq 185 \; {\rm MeV/c}$) and for the high partial waves at
low energy. The method allows for fixing two of the constants in the
subthreshold expansion for $\bar{D}^+$ in powers of $\nu^2$ and $t$
\begin{eqnarray}
\bar{D}^+=d_{00}^++d_{10}^+\nu^2+d_{01}^+t+d_{20}^+\nu^4+d_{11}^+\nu^2t+...,
\nonumber
\end{eqnarray}
where
\begin{eqnarray}
d_{00}^+=\bar{D}^+(0), \; \; \; \; d_{01}^+=\bar{E}^+(0). \nonumber
\end{eqnarray}
The curvature term $\Delta_D$ is defined by
\begin{eqnarray}
\Sigma=F_\pi^2(d_{00}^++2M_\pi^2d_{01}^+)+\Delta_D \equiv \Sigma_d + \Delta_D,
\nonumber
\end{eqnarray}
where $\Delta_D$ is dominated by the $\pi \pi$ cut giving \cite{gls1}
\begin{eqnarray}
\Delta_D=11.9 \pm 0.6 \; {\rm MeV}.
\nonumber
\end{eqnarray}
$\Sigma_d$ is a sensitive quantity as is demonstrated by the numerical values for the
two solutions, A and B, of ref. \cite{gls2}, where $\Sigma_d = 48 - 50$ MeV with an
error of about 10 MeV.
Now the question is how the value for $\Sigma$ would change, if amplitudes
based on modern meson factory data would be used as input instead of the Karlsruhe
amplitudes where input mostly consisted of data before the meson factory era.
With the VPI/GWU input (SM99 and SM01) \cite{arndt} results typically in the range
\begin{eqnarray}
\Sigma_d=([-80. \, {\rm to} \, -77.] + [146. \, {\rm to} \,  157.]) \; {\rm MeV} = 65. \,  {\rm to}\, 80. 
\; {\rm MeV}
\nonumber
\end{eqnarray}
follow, i.e. a considerably larger value than the Karlsruhe input would give.
The corresponding $\pi^-$p scattering length
\begin{eqnarray}
a_{\pi^- p}=0.0857 - 0.0899 \; M_\pi^{-1} \nonumber
\end{eqnarray}
is to be compared with the experimental value $0.0883 \pm 0.0008 \; M_\pi^{-1}$
\cite{sch}.

\section{Summary}

In a lattice calculation the value for $\sigma$ can be obtained from the quark
mass expansion of the nucleon mass by making use of the Feynman-Hellmann theorem
as given above. A new development has recently been the inclusion of dynamical
quarks \cite{sesam} giving $\sigma = 18 \pm 5 \; {\rm MeV}$. In general, there is,
however, the problem
that the value for $m_q$ is still quite large and the extrapolation to
small quark mass values is uncertain.

The nucleon mass in full (two-flavour) QCD as a function of the pion
mass has been calculated by UKQCD \cite{ukqcd} and CP-PACS \cite{cppacs} 
collaborations. The values so found have been fitted with a ChPT-inspired 
expression \cite{lwb}
\begin{eqnarray}
m_N = \alpha + \beta M_\pi^2 + \sigma_{NN}(M_\pi, \Lambda) + 
    \sigma_{N\Delta}(M_\pi, \Lambda) \nonumber
\end{eqnarray}
for the quark mass dependence of the nucleon mass leading to the result
$\sigma = 45 - 55 \; {\rm MeV}$.
The functions $\sigma_{NN}$ and $\sigma_{\Delta N}$ are due to the nucleon
self-energy diagrams with an intermediate nucleon and delta respectively.

Promising steps have been made in the lattice frontier, but still more work
is needed. For the phenomenological part questions remain. It turns out that
$\Sigma$ is a quite sensitive quantity and, therefore, requirements of
consistency of the low-energy data and analysis are of particular importance.
E.g., $\Sigma_d$ is sensitive to the high partial waves at low energy.
An additional problem is the question of electromagnetic corrections
close to the physical threshold, see \cite{lr,fm}.

\acknowledgments{I wish to thank A.M. Green for useful remarks on the manuscript.
Financial support of the Academy of Finland grant 47678, the TMR
EC-contract CT980169 and the Magnus Ehrnrooth Foundation is gratefully acknowledged.}


\end{document}